\documentclass[acmsmall,screen,nonacm]{acmart}
\def\BibTeX{{\rm B\kern-.05em{\sc i\kern-.025em b}\kern-.08emT\kern-.1667em\lower.7ex\hbox{E}\kern-.125emX}}
    
\usepackage{silence}
\ActivateWarningFilters[todo]
\usepackage{multirow}

\usepackage{pifont}
\usepackage{subfig}
\begin{document}

%
\title{Design of Energy Harvesting based Hardware for IoT Applications}

%
\author{SatyaJaswanth Badri}
\email{2018CSZ0002@iitrpr.ac.in}
\author{Mukesh Saini}
\email{mukesh@iitrpr.ac.in}

\author{Neeraj Goel}
\email{neeraj@iitrpr.ac.in}

\affiliation{%
  \institution{Indian Institute of Technology Ropar}
  \streetaddress{S.Ramanujan Block, IIT Ropar Main Campus}
  \city{Ropar}
  \state{Punjab, India}
  \postcode{140001}
}

%
\renewcommand{\shortauthors}{S.J Badri, et al.}

%
\begin{abstract}
Internet of Things (IoT) devices are rapidly expanding in many areas, including deep mines, space, industrial environments, and health monitoring systems. Most of the sensors and actuators are battery-powered, and these batteries have a finite lifespan. Maintaining and replacing these many batteries increases the maintenance cost of IoT systems and causes massive environmental damage. Energy-harvesting devices (EHDs) are the alternative and promising solution for these battery-operated IoT devices. These EHDs collect energy from the environment and use it for daily computations, like collecting and processing data from the sensors and actuators. Using EHDs in IoT reduces overall maintenance costs and makes the IoT system energy-sufficient. However, energy availability from these EHDs is unpredictable, resulting in frequent power failures.

Most of these devices use volatile memories as storage elements, implying that all collected data and decisions made by the IoT devices are lost during frequent power failures, resulting in two possible overheads. First, the IoT device must execute the application from the beginning whenever power comes back. Second, IoT devices may make wrong decisions by considering incomplete data, i.e., data-inconsistency issues. To address these two challenges, a computing model is required that backs up the collected data during power failures and restores it for later computations; this type of computing is defined as intermittent computing. However, this computing model doesn't work with conventional processors or memories. Non-volatile memory and processors are required to design a battery-less IoT device that supports intermittent computing.
\end{abstract}

%
%

%

%

%
\maketitle

\section{Introduction}

The Internet of Things (IoT) created many fascinating applications allowing users to interact with numerous sensors and actuators. In many areas, such as deep mines, health monitoring systems, industrial applications, and many other fields, IoT helps to perform tasks that humans cannot do. For example, IoT devices were used in the healthcare community to monitor glucose, heart rate, and stress levels \cite{1,2}. Recent advancements show that IoT devices assist in robotic surgeries and effectively keep pacemakers near the patient's heart \cite{pace}.

According to a study, billions of these sensors and actuators will be used in real-time IoT environments by the end of 2050 \cite{big}. Most of these IoT devices were powered by standard or rechargeable batteries, which have a limited lifespan \cite{lifetime}. Monitoring or replacing these many batteries increases maintenance costs. Disposing of these many battery-powered IoT devices harms the environment and reduces the device's lifespan.

The alternative and promising technology that assists in replacing battery technology is Energy-Harvesting devices (EHDs). These EHDs collect energy from our surroundings \cite{21, e1}, such as solar, piezo-electric, wind, motion, breathing, etc. Solar voltaic cells, for example, extract energy from the sun \cite{e1}. Using these EHDs in an IoT environment extends the life of the IoT devices. At the same time, the cost of maintaining or charging these voltaic cells is significantly lower and more cost-effective. Figure 1 shows the various energy-harvesting sources that are available in our surroundings.

\begin{figure}[htb]
\centering
\includegraphics[width= 0.78\linewidth]{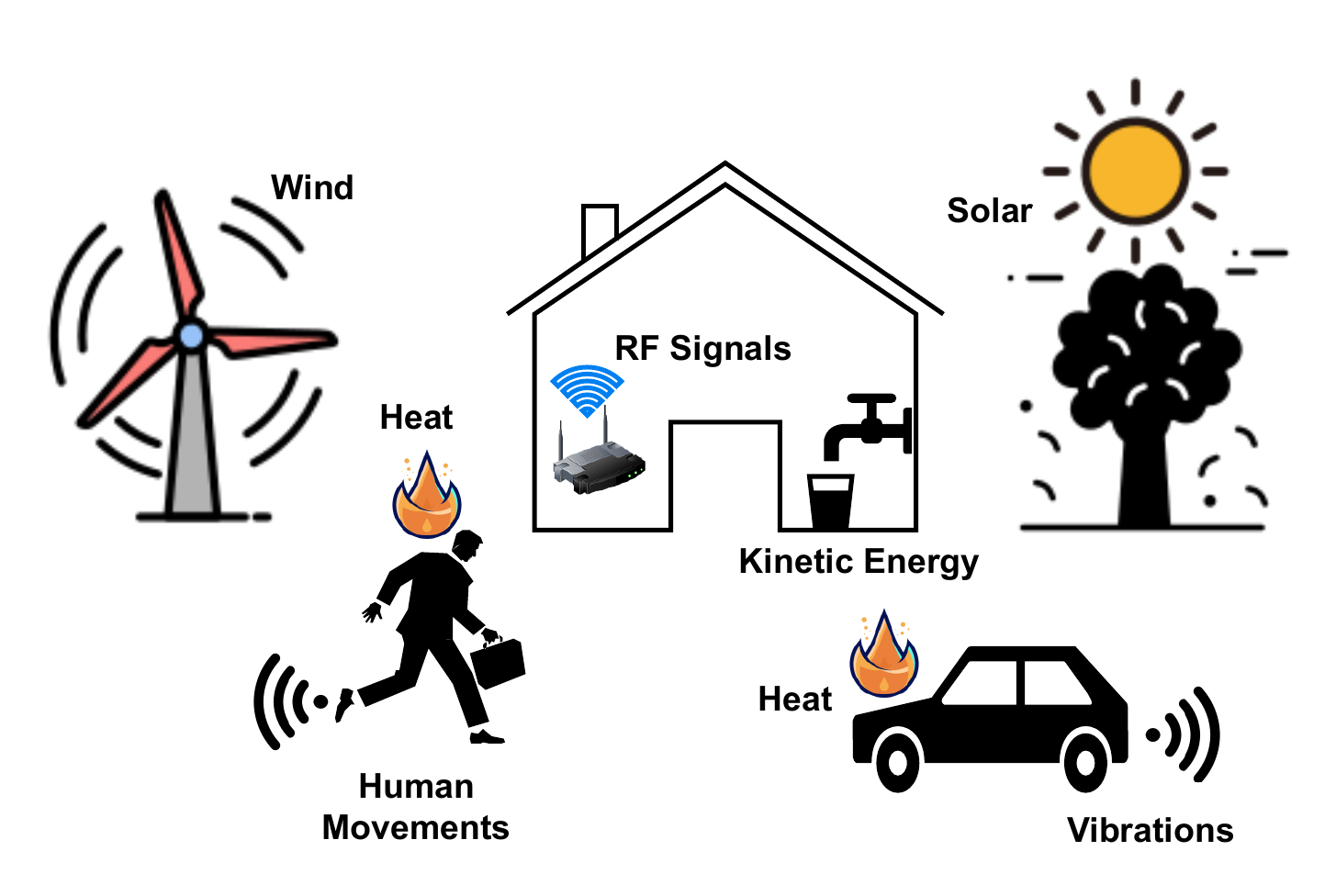}
\textbf{ \caption{Different Energy Harvesting Resources Available in the Surroundings}}
 \label{fig1}
 \end{figure}

 These EHDs collect energy from their surroundings and store it in a capacitor, where the stored energy is used by IoT devices to perform necessary computations. Incorporating EHDs and a capacitor with the IoT devices makes them battery-free devices \cite{capac}. However, the energy available from these EHDs is unstable, which may result in frequent power failures. During power failures, the execution of these IoT applications may become irregular, resulting in inconsistent output. As a result, a computing model is required that enables IoT devices to backup and restore all the computed results that are executed during these frequent power failures. This computing model is known as intermittent computing \cite{8,9,10}. This chapter discusses the challenges associated with these devices and the hardware required for these battery-free IoT devices that use harvested energy for executing the IoT applications.

\section{Energy Harvesting in IoT devices \& Environment}
This section discusses the available energy sources in our surroundings and how these EHDs benefit IoT devices and the environment. As shown in Figure 1, there are four to five energy sources that are available in our surroundings, including solar energy from the sun, mechanical energy from a nearby windmill, RF energy from a Wi-Fi router, kinetic(mechanical) energy from water flow, vibration(mechanical) energy from the road, and thermal energy from the heat produced by the human body and the car. Human breathing and body movements can also generate a small amount of energy.

Deploying EHDs for each energy source may accumulate a good amount of energy, which can be helpful in executing IoT applications. EHDs such as solar voltaic cells and PV panels harvest solar energy, which extracts energy in the range of $15-100 mW/cm^2$. The primary benefit of solar energy is its consistent behavior; during the day, it harvests a large amount of energy from its surroundings. The only drawbacks are its unpredictable behavior and deployment constraints. EHDs like anemometers are used to harvest wind energy, which extracts approximately $1200 mWh/day$. The main advantage of wind energy is its ease of deployment, and it can use in open areas. The only disadvantage for these EHDs is maintenance costs.

EHDs, similar to piezoelectric materials, collect energy from human movements, footfalls, and car vibrations on the road, extracting approximately $2.1-5 W$ for human finger movements/footfalls and $200 \mu W/cm^2$ for human motion. Piezoelectric materials are fully controllable and produce energy based on our needs. However, piezoelectric materials extract the least amount of energy. EHD, like a ratchet flywheel, collects energy from human breathing, extracting $0.42W $ of energy approximately. This energy-harvesting source is advantageous because it is always available in our surroundings but produces little energy.

For extracting thermal energy (body heat, car engine heat), EHDs, such as thermo-couple batteries, collect approximately $50 mW/cm^2$. The main advantage of these EHDs is that they are more reliable, require less maintenance, and last longer. However, the energy conversion is very low and insufficient for an IoT environment without additional support. EHDs such as rectennas extract RF energy, which collects about $1 W/cm^2$. The main advantage of these EHDs is their mobility and high energy density. However, these EHDs produce radioactivity, which may be more dangerous to people nearby.

\subsection{Internals of Energy-Harvesting-based IoT devices}
Many of the EHDs are unpredictable and uncontrollable. Using energy-harvesting resources in an IoT environment needs a proper energy management scheme. Figure 2 shows the three main components of an energy-harvesting system  \cite{e1,e2,e3,e4,e5}.

\begin{enumerate}
    \item \textbf{Energy-Harvesting Resources:} Resources that gather energy from their surroundings. 
    \item \textbf{Energy-Monitoring System:} Once the energy is extracted, a storage mechanism is required to store and manage it.
    \item \textbf{IoT devices \& Environment:} Effective utilization of stored energy is needed for these IoT devices \& the environment.
\end{enumerate}

\begin{figure}[htb]
  \includegraphics[width= 1\linewidth]{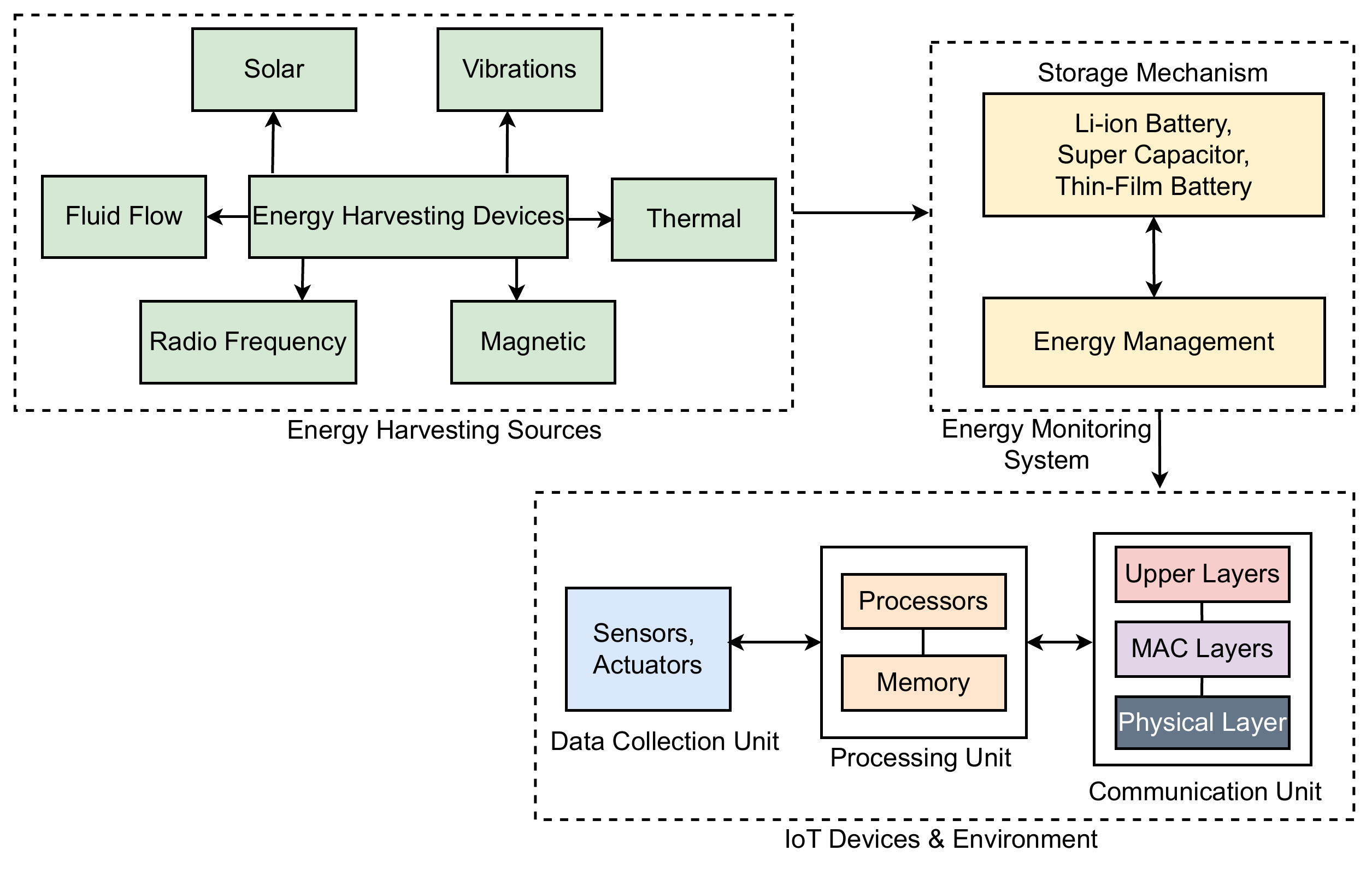}
  \label{fig21}
\textbf{ \caption{An Overview of Energy-Harvesting-based Architecture for IoT Environment}}
 \end{figure}

After gathering energy from the environment, efficient mechanisms for storing the collected energy are required. Rechargeable batteries, supercapacitors, and thin-film batteries, among others, store enough energy within these materials. Supercapacitors/thin-film batteries experience significant energy leakage during charging and discharging cycles. As a result, management techniques are also required to efficiently use accumulated energy by allocating sufficient energy to IoT sensors and actuators based on their needs. An efficient scheme or technique is required to monitor these leaks and implement a near-zero leakage policy. Figure 2 shows an overview of the energy-harvesting architecture used in the IoT environment.

The energy-monitoring system distributes energy to all IoT devices deployed in the field. The majority of IoT devices and environments are composed of three subsystems:

\begin{enumerate}
    \item The data collection unit contains many sensors and actuators that collect data based on the IoT application.
    \item To store the computed results, a processing unit with memory and processors is required to make the correct decision based on the data collected.
    \item Once a decision is made in a specific situation, the final decision must be communicated to the other IoT device or end user via the communication unit.
\end{enumerate}

\subsection{Challenges associated with these energy-harvesting-based IoT devices}
This section discusses the challenges associated with the energy-harvesting-based IoT architecture shown in Figure 2.

\begin{itemize}
    
    \item \textit{\textbf{Challenge 1:}} Energy availability is unpredictable and irregular. These EHDs have no control over when and how much energy they collect.
    \item \textit{\textbf{Challenge 2:}} In the IoT environment, Issue-1 causes frequent power failures. Because of these power failures, the execution model of the IoT application becomes irregular.
    \item \textit{\textbf{Challenge 3:}} Figure 2 shows how supercapacitors are used as energy storage mechanisms in an energy monitoring system. Once capacitors are connected to IoT devices, the size of the capacitors is fixed, as is the amount of energy stored in these capacitors. As a result, this constant amount of energy should be used efficiently for IoT computations.
    \item \textit{\textbf{Challenge 4:}} There will be energy leakage from these capacitors/batteries and an effective method is needed to monitor these energy leakage issues.
    \item \textit{\textbf{Challenge 5:}} Traditional processors and memory storage models are volatile in today's world. In these energy-harvesting-based IoT architectures, data collected by IoT sensors and computation results may be lost during power failures. When power is restored, the IoT application must start over, consuming more energy by repeating the same procedures.
\end{itemize}

This chapter contributes to the discussion of the major problems related to the energy-harvesting-based IoT architecture. This chapter also discusses potential hardware for addressing the challenges listed above.

\section{Intermittent Computing}
When there is enough harvested energy in a capacitor and the energy harvester directly provides enough energy, the IoT application is run as usual, using the energy directly from the harvester source. When the harvester does not provide energy directly from the source, the IoT device must rely on energy from the capacitor to perform essential tasks. Figure 3 shows enough solar energy during the day for the IoT device to run the application without intervention around 01:00 PM. The IoT device must use the capacitor's energy to complete important tasks before turning off in the evening, around 05:00 PM.

\begin{figure}[htb]
  \includegraphics[width= 1\linewidth]{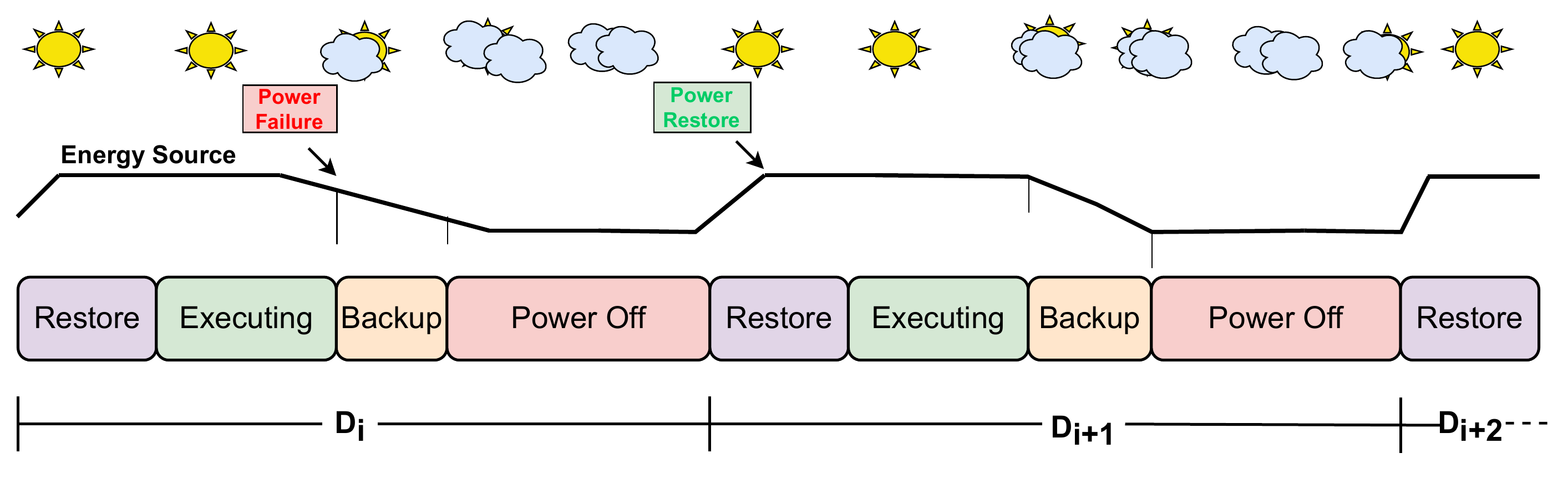}
  \label{fig3}
 \textbf{\caption{An Overview of Intermittent Computing for Solar-based Harvesting Systems }}
 \end{figure}

Because energy is not always available to harvest, execution in such devices is intermittent, resulting in power failures in the IoT environment \cite{8,9,10}. Even when energy is readily available, it takes time to accumulate enough energy to perform valuable work. Incorporating new memory technologies and additional procedures into the execution and memory model of a conventional processor is required to design an intermittently aware design.  

\subsection{Memory Model for Intermittently Powered Devices}
\begin{figure}[htb]
\centering
\includegraphics[width= 1\linewidth]{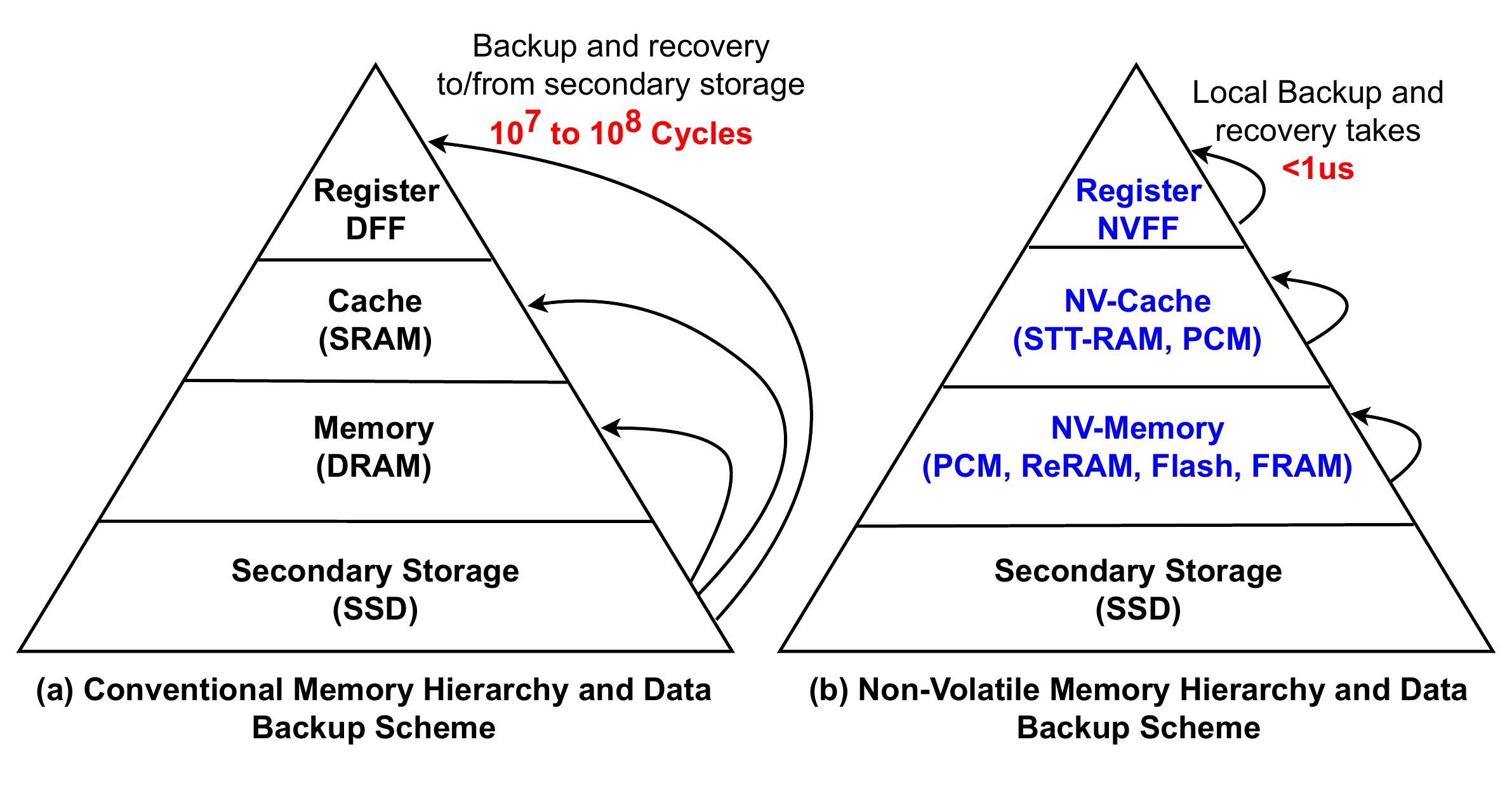}
 \caption{\textbf{Differences between Conventional and Non-Volatile Memory Models}}
 \label{fig10}
 \end{figure}

The hardware of an intermittently functioning device may include general-purpose computing units such as a CPU or a microcontroller unit (MCU), a group of sensors, and one or more radios to communicate with the sensor array. Almost all of these devices use a volatile memory model. Figure \ref{fig10} (a) shows the conventional memory hierarchy, which includes the register file, caches, main memory, and secondary storage. All other memories are volatile except for secondary storage and can hold data only whenever power is available. As illustrated in Figure 3, the IoT device will shut down when no energy is available from the harvester source, i.e., at night (after 7:00 PM); during this time, data stored in registers, caches, and main memory is lost.

There are two alternatives to keep the volatile contents safe. One approach is to save all computed results and decisions made during the execution phase to secondary storage before the MCU enters the power OFF state. The second approach is to restart the IoT application whenever the energy harvester provides enough energy to the MCU. Both alternatives are inefficient because backup/recovery to/from secondary storage and re-executing the same application take more time and energy. As a result, it is necessary to have a memory model that can store the memory contents when the energy source is unavailable.

Non-volatile memory (NVM) is a relatively new memory technology that can retain the system state while consuming no power. NVM technologies are being developed to overcome the drawbacks of volatile memory technologies. Flash, spin transfer torque RAM (STT-RAM), phase-change memory (PCM), resistive RAM (ReRAM), and ferroelectric RAM (FRAM) are examples of emerging NVM technologies. Because of their exhibited physical properties, NVMs have the potential to consume very little power while providing significantly greater density than conventional memory technologies. A standard SRAM cell, for example, has a size of $125-200F^2$, and a PCM and Flash cell have sizes of $4-12F^2$ and $4-6F^2$, accordingly, where F refers to the lowest lithographic dimensions that range in a specific technology node. Because of their advantages, NVMs have become more common in products. Flash memory, for example, is used as a cache in Intel TurboMemory.

These NVMs, however, have some limitations. NVMs, for example, have a higher latency and consume more energy than volatile memory technology. Write endurance is the property that determines how many writes a memory block can withstand before it becomes ineffective. NVMs have significantly lower write endurance than traditional memory technologies. Table 1 provides detailed comparisons of various properties with various memory technologies. Access granularity is defined as the minimum size of data read/written in each access. Furthermore, they can store data for many years without requiring standby power under regular circumstances. As shown in Table 1, the common insight from all NVM technologies is that write latency/energy is greater than read latency/energy \cite{s1,s3,s5,s6,s7}.

\begin{table}[htp]
\centering
\textbf{\caption{Comparisons between different NVM Technologies for different Features}}
\label{tab1}
\resizebox{\columnwidth}{!}{%
\begin{tabular}{|l|l|l|l|l|l|l|l|l|}
\hline
\textbf{Property} &
  \textbf{SRAM} &
  \textbf{DRAM} &
  \textbf{HDD} &
  \textbf{SLC Flash} &
  \textbf{PCM} &
  \textbf{STT-RAM} &
  \textbf{ReRAM} &
  \textbf{FRAM} \\ \hline
\textbf{Cell Size ($F^2$)} &
  $120-200$ &
  $6-12$ &
  NA &
  $4-6$ &
  $4-12$ &
  $6-50$ &
  $4-10$ &
  $12-15$ \\ \hline
\textbf{Read Latency} &
  $\sim$1 ns &
  $\sim$110 ns &
  5 ms &
  $25 \mu s$ &
  50 ns &
  <10 ns &
  <10 ns &
  50 ns \\ \hline
\textbf{Write Latency} &
  $\sim$1 ns &
  $\sim$110 ns &
  5 ms &
  $500 \mu s$  &
  500 ns &
  10 ns &
  <10 ns &
  50 ns \\ \hline
\textbf{\begin{tabular}[c]{@{}l@{}}Write Energy\\ (J/bit)\end{tabular}} &
  $\sim$ $10^{-15}$ &
  $\sim$ $10^{-14}$ &
  $\sim$ $10^{-14}$ &
  $\sim$ $10^{-9}$ &
  $\sim$ $10^{-11}$ &
  $\sim$ $10^{-13}$ &
  $\sim$ $10^{-13}$ &
  $\sim$ $10^{-12}$ \\ \hline
\textbf{Leakage Power} &
  High &
  Medium &
  Medium &
  Low &
  Low &
  Low &
  Low &
  Low \\ \hline
\textbf{Erase Latency} &
  NA &
  NA &
  NA &
  2 ms &
  NA &
  NA &
  NA &
  NA \\ \hline
\textbf{\begin{tabular}[c]{@{}l@{}}Access Granularity\\ (B)\end{tabular}} &
  64 &
  64 &
  512 &
  4K &
  64 &
  64 &
  64 &
  64 \\ \hline
\textbf{Endurance} &
> $10^{16}$ &
  > $10^{16}$ &
  > $10^{15}$ &
  $10^4 - 10^5$ &
 $ 10^8 - 10^{15}$ &
  >$10^{15}$ &
  $10^8 - 10^{12}$ &
  $10^{10} - 10^{12}$ \\ \hline
\textbf{Standby Power} &
  $0.6 - 1.2 W$ &
  Refresh Power &
 $ 1 - 2 W $ &
  0 &
  0 &
  0 &
  $0$ &
  $0$ \\ \hline
\end{tabular}%
}
\end{table}

In terms of characteristic features, STT-RAM outperforms all other NVM technologies. Table 1 shows that STT-RAM outperforms other NVMs in terms of write endurance, latency, and energy consumption. As a result, STT-RAM is a promising candidate for cache, main memory, and scratch-pad memory. However, because STT-RAM is more expensive than other NVMs, it is unsuitable for use at the main memory level. PCM is the next better NVM technology after STT-RAM because its size and write endurance are better than other NVMs. As a result, PCM is a promising candidate for use in cache and main memory. However, because STT-RAM has a longer lifespan than PCM, it is unsuitable for use at the cache level. PCM is the better candidate for main memory. ReRAM and FRAM share some characteristics; both were promising candidates for main memory. ReRAM, on the other hand, has lower latency and energy consumption than FRAM. For example, the MSP430FR6989 is a recent TI-based microcontroller with 2KB SRAM and 128KB FRAM at the main memory level. 

Incorporating NVMs at each level preserves the data in these IoT devices from frequent power failures, switching the traditional processor into a non-volatile processor. (NVP). Figure \ref{fig10} (b) shows the memory hierarchy of non-volatile flip-flops, non-volatile caches (STT-RAM/PCM), and non-volatile main memory (PCM, ReRAM, and FRAM). Thus, replacing the non-volatile memory model with volatile memory helps in intermittent computing and reduces the time and energy required to backup/retrieve volatile contents.

\subsubsection{Designing NVM-based Processor for Intermittently Powered IoT Devices}
A non-volatile processor (NVP) is designed by replacing volatile memory with NVM at each level. Figure 5 shows two distinct architectures that demonstrate the differences between pure NVM architecture and hybrid NVM architecture. There is a pure NVM technology at each level in architecture-1, particularly STT-RAM at both the L1 and last-level cache (LLC) levels and PCM at the main memory level. In architecture-2, hybrid NVM technology is used at each level, with SRAM+STT-RAM at the L1 and LLC levels and SRAM+PCM at the main memory level \cite{xie,msp,n2,backup,n1}. 

\begin{figure}[htb]
\centering
  \includegraphics[width= 0.7\linewidth]{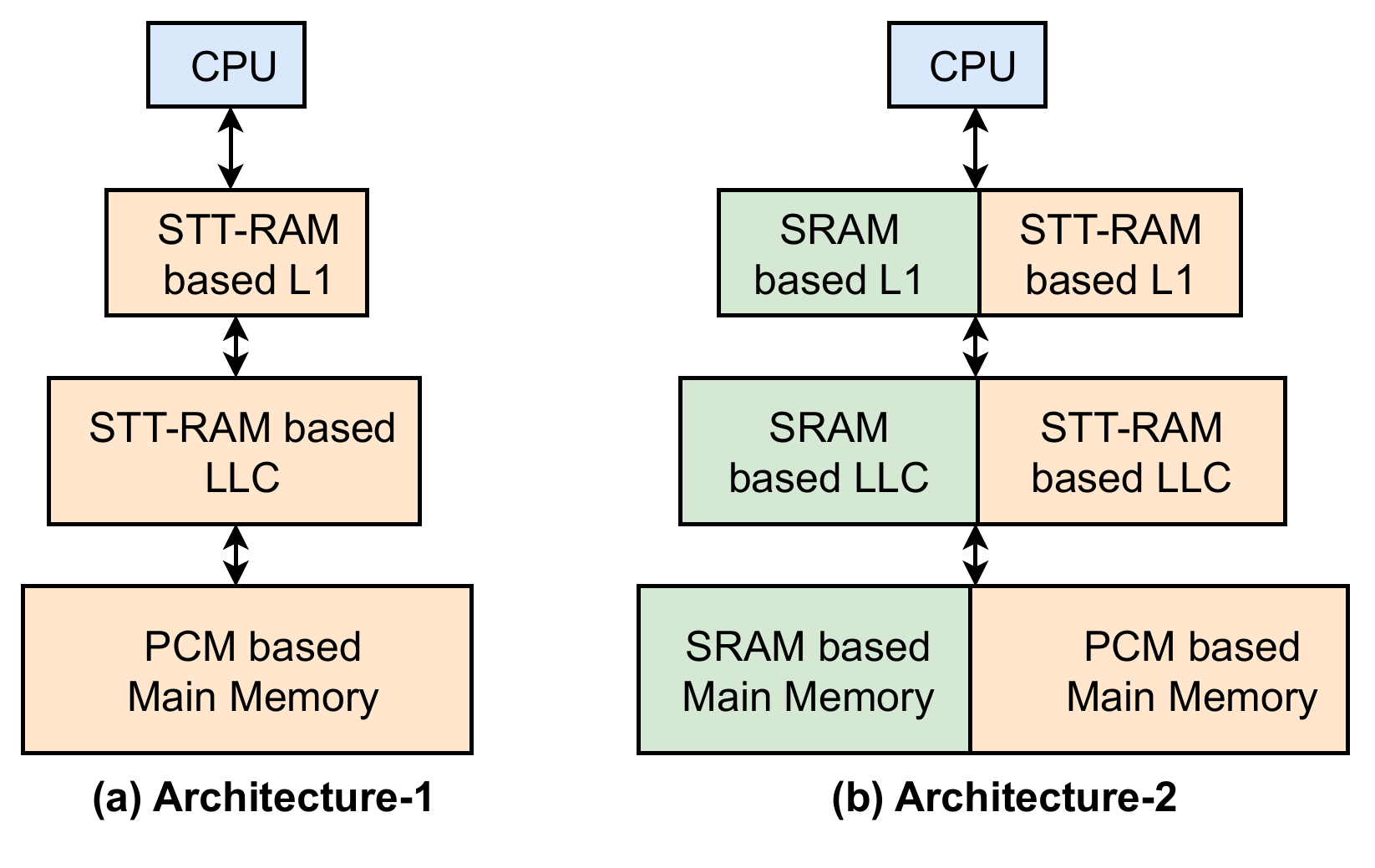}
 \textbf{\caption{(a) Pure-NVM based architecture, STT-RAM at cache levels \& PCM at main memory level and (b) Hybrid-NVM based architecture, SRAM+STT-RAM at cache levels \& SRAM+PCM at main memory level }}
 \label{fig32}
 \end{figure} 

The main advantage of hybrid architecture over pure NVM architecture is that it takes advantage of both SRAM and NVM, i.e., performance benefits from SRAM and non-volatility and density benefits from NVM. Figure 5 (b) shows a design that allows you to experiment with various combinations. As needed, the designer must select hybrid NVM and pure NVM architectures. The main disadvantage of using hybrid NVM over pure NVM architecture is that the volatile contents in the hybrid architecture must be stored in NVM during frequent power failures, which increases backup time and energy.

 \subsection{Execution Model for Intermittently Powered Devices}
Regardless of several significant differences between the intermittent execution model and the conventional embedded execution model, designers of today's intermittently powered devices use a standard, C-like embedded computing abstraction. The application on an intermittently powered device runs until the device's energy has been drained. When the application's energy is restored, it resumes the execution from a specific point in its execution history, such as the start of the main() function or a safe point. The primary distinction between conventional and intermittent execution models is that a normally executing program is expected to run until it is completed. In contrast, an intermittent execution model must complete the program execution despite multiple power interruptions. Various system components, such as languages, application run time behavior, and program semantics, must be modified to create an intermittence-aware design.

Among all of these changes, we highlighted three significant changes required in the application execution flow for the intermittence-aware design. 

\textbf{Checkpoint():} When a checkpoint procedure is detected, all volatile contents are copied to NVMs in order to preserve the system state. The existing literature in this area focuses on when and where to place a checkpoint. Recent research proposes two methods for determining when to perform this checkpoint procedure.  

\begin{enumerate}
    \item To monitor the energy source and capacitor, specially designed hardware is required. When it falls below a certain threshold, the system sends an interrupt, which stops the application and starts the backup() procedure. Thus, checkpoints can occur at any time. 
    \item Instead of maintaining hardware to monitor the energy requirements, existing solutions track changes in the application state. There are solutions that determine the variation at checkpoint time, either through hash comparisons or by comparing main memory word-by-word with the most recent checkpoint data, which is already in NVM. When there are many changes to initiate a backup() procedure, the system issues an interrupt.
\end{enumerate}

\textbf{Backup():} Whenever a backup() procedure is initiated, it copies the volatile contents to NVM, which means that it reads the contents from volatile memory technology and writes them to NVM technology.

\textbf{Restore():} Whenever a restore() procedure is initiated, it copies the backed-up contents from NVM to volatile memory, which means it reads the contents from NVM technology and writes them to volatile memory technology.

Adding additional procedures such as checkpoint(), backup(), and restore() to the conventional execution supports intermittent computing, which also completes Figure 3. However, these additional procedures may incur additional costs. To reduce additional overheads in the intermittent execution model, efficient checkpointing, backup, and rollback policies are required. Both the execution and memory models support intermittently powered devices and computing for these devices.
 
\subsection{Challenges associated with these Intermittently Powered IoT devices}
Intermittent computing introduces a number of challenges. These challenges demonstrate how the system becomes inconsistent and fails to progress, assisting an individual in developing an efficient intermittent-aware design by dealing with these challenges properly.

\subsubsection{Application Progress}
If the application's system state is not backed up completely across all power failures, the application will re-execute from the beginning, i.e., from the main() function \cite{in1,8,9,10}. Figure 6 (b) shows how the execution progress is affected by frequent power failures. In Figure 6 (a), there is a backup() after the first couple of instructions, and a power failure occurs after processing(result). When power is restored, execution begins with sensing() rather than main(), saving energy and time. However, when a power failure occurs again after processing (result), the application enters an infinite loop, as illustrated in Figure 6 (b). It never advances the execution of that instruction, and as a result, the application never completes its execution and returns no results.

\begin{figure}[htb]
\centering
  \includegraphics[width= 0.776\linewidth]{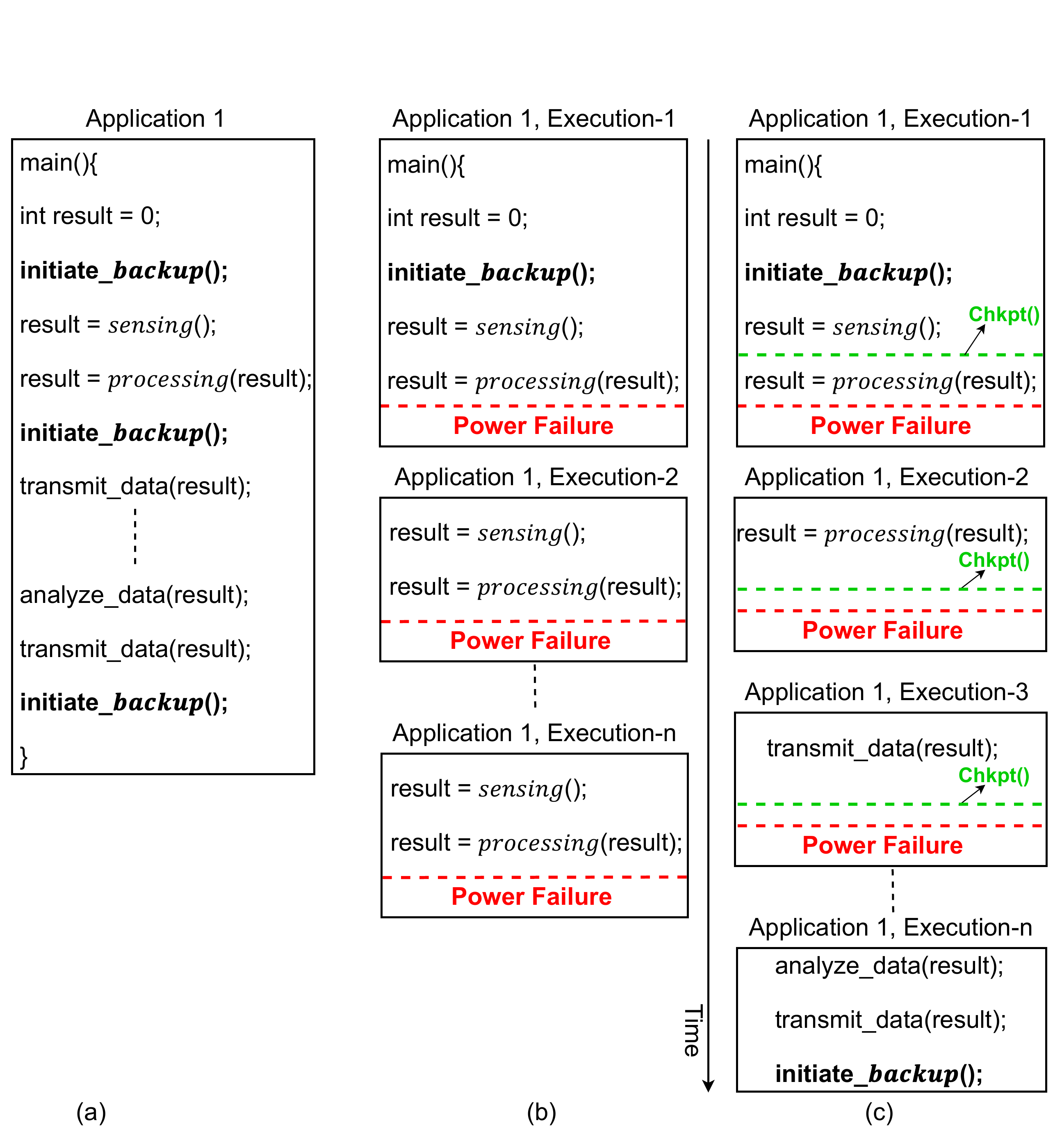}
 \textbf{\caption{A detailed Example of Irregular Execution Progress during Intermittent Power Supply (a) Application-1 without any Power failures, (b) Execution of Application-1 during frequent Power failures, and (c) Execution of Application-1 during frequent Power failures using the checkpointing technique. }}
 \label{fig31}
 \end{figure} 

Specific checkpointing techniques must be incorporated into the application in order to make consistent progress. Figure 6 (c) illustrates how inserting a checkpoint after each instruction improves execution progress over the previous approach (Figure 6 (b)). The programmer must ensure that each checkpoint consumes one cycle of capacitor energy and that the distance between checkpoints is acceptable to charge the capacitor sufficiently.

\subsubsection{Memory Consistency}
Most intermittently powered MCUs can have a mix of volatile and NVMs; the MSP430FR6989, for example, has SRAM and FRAM. A naive design of volatile memory and NVMs could lead to inconsistency in both memories \cite{in1,8,9,10}. Figure 7 shows the application performing two tasks: incrementing an NVM variable, saving variable data into an array1, and adding array elements, saving the final value in the variable, where all array elements and results are stored in NVM. 

\begin{figure}[htbp]
 \centering
  \includegraphics[width= 0.55\linewidth]{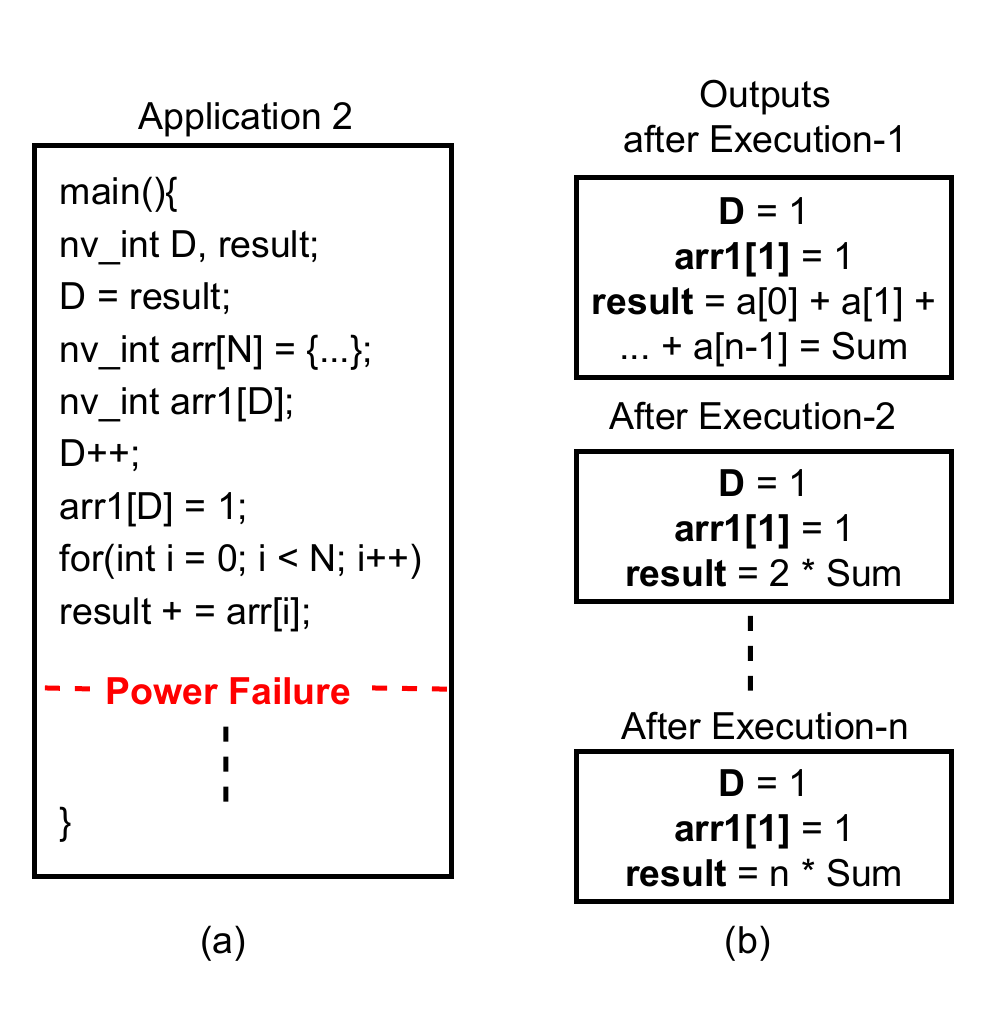}
 \textbf{\caption{A detailed Example of Memory Inconsistency behavior during Intermittent Power Supply (a) Application-2 without any Power failures and (b) Outputs after Executing Application-2 during frequent Power failures.}}
 \label{fig311}
 \end{figure}
 
Assume there is a power failure between the increment operation and saving the data into the array. In that case, the application's entire system state is lost, but not the data stored in the NVM, such as ``D,'' ``arr1[],'' and ``result,'' as shown in Figure 7. When power is restored, the application begins again, unaware of the fact that the data in NVM has already been computed/incremented. As a result, in execution-2, the application increments the variable ``D'' once more and computes the variable ``result'' using the incorrectly stored array values, as illustrated in Figure 7 (b). During frequent power failures, these memory inconsistencies increase, as illustrated in Figure 7 (b). If the run-time system permits certain parts of the program to be re-executed, it must also monitor and recover memory addresses that could cause inconsistencies.

\subsubsection{Preserving Program Semantics}
Even if an application has well-suited checkpoints and a system that maintains a consistent state between NVM and volatile memory, it may perform differently than the designer expected. When the capacitor discharges and a power failure occurs, the energy harvesting device is turned off for a defined period of time, and all peripherals and their system states are cleared. When compared to continuously powered devices, this behavior may violate the designer's assumptions about the atomicity of operations and the timeliness of data.

\begin{enumerate}
    \item \textbf{Atomicity:} Certain code regions must execute sequentially (with no checkpoints in the middle), ensuring the application runs correctly. Before reading from a sensor device, for example, the device driver checks that the sensing device is turned on (enough energy is provided to the sensor) and that the bus is ready to use because many sensor devices use this bus for communication with other devices. Assume there is a checkpoint between these program assertions and the sensor read function, and a power failure occurs. When power is restored, the application returns to the previously saved checkpoint and executes the sensor's read function directly without performing the initial checks. Designers should be able to specify which atomic regions must be completed within a single energy peak.

     \item \textbf{Timeliness:} Some information loses value over time. Because the device may remain turned off for a prolonged period of time if the power goes away, placing checkpoints between the gathering and utilizing data restricts its usefulness.

Consider an application that monitors a specific system's temperature and sends a radio signal alert indicating whether the temperature is appropriate or exceeds a certain threshold. The application takes temperature readings but terminates before they are completed. When the device is turned off, the temperature changes dramatically. Whenever the application is restarted, the device continues to process the stale data and incorrectly reports that the system's temperature readings are still acceptable. Programmers must be able to communicate which events must occur in a timely fashion.
\end{enumerate}

\section{Future Research Directions}
Intermittent computing in energy-harvesting-based IoT devices is a new research area that has the potential to design battery-free devices. 

\textbf{Programming Intermittent Devices:} Despite the growing popularity of intermittent devices, recent approaches to using them have several significant drawbacks: (a) the developer's effort in defining the tasks that IoT devices must perform has been increased, (b) deciding the optimal size of each task, (c) additional Overheads are due to run time exceptions and memory models, and (d) inaccurate modeling of application-level characteristics (e.g., I/O atomicity).

The limitations motivate more investigations to address the challenges mentioned above by developing new programming models that reduce overheads and the developer's burden while maintaining developer control over atomicity.

\textbf{Designing Distributed environment of intermittent devices:} A communication platform is required for distributed intermittent computing devices to communicate with one another. The time between each communication consumes less energy than the fixed communication span between these intermittent computing devices compared to traditional communication platforms. Synchronizing the collected data from these devices is an unresolved issue, and monitoring the differences between synchronized and unsynchronized data from these devices is also an open research problem. 

The development of distributed systems that use intermittent devices allows exceptional battery-free applications. The lack of programming tools \& environment, programming language specifications, memory \& execution model abstraction increases the complexity of designing an accurate, efficient, distributed intermittent computing system. Recent measures have focused on intermittent distributed shared memory models and simulator-based support.

\textbf{Effective use of NVMs:} Incorporating NVMs into intermittent systems consumes more energy. So, when including NVM, efficient management and prediction policies are required to reduce the overheads caused by NVMs and to use NVMs efficiently. When using NVMs for hybrid caches/main memories, an optimal placement policy that predicts which block should be placed in which memory region is a crucial research objective. Recent research has centered on this objective, proposing novel placement and prediction policies.

\textbf{Efficient Checkpointing Policies:} Checkpointing techniques are required to keep the execution progress consistent. In an intermittent application programmer inserts the checkpoints wherever it is needed. More checkpoints in an application increase overheads due to unnecessary backup/restore to/from NVMs and an increasing number of NVM accesses. As a result, the checkpointing overhead consumes more energy than standard backup/restore procedures. Determining when to checkpoint and which parts of an application to backup is an unexplored research area. Reducing the number of unnecessary checkpoints helps to reduce the time and energy required to backup/restore volatile contents during a power failure.

\section{Conclusions} \label{p6}
Battery-free devices are required for today's IoT applications. Integrating EHDs into an IoT environment introduces new challenges. In order to address these challenges and gain support for intermittent computing, there is necessary to change the conventional execution and memory model. The NVM model, which replaces the traditional volatile memory model, assists in support of intermittent computing. NVMs can store data even during a power failure, making them useful as a backup/restore region for intermittent computing. Many new NVM technologies have been proposed and are now on the market. However, NVMs alone are insufficient for intermittent computing and challenges due to unstable energy availability from harvesting resources. The intermittent execution model includes different application procedures that handle when and what to backup/restore to/from NVMs during frequent power failures.

This chapter covers all aspects of intermittent computing for EHD-based IoT devices. This chapter addresses all the modifications and hardware required to replace existing models and energy-harvesting-based architectures in the IoT environment. This chapter discusses how to turn battery-powered IoT devices into batteryless IoT devices. 

\section*{Acknowledgement}
This work is supported by the grant received from the Department of Science and Technology (DST), Govt. of India, for the Technology Innovation Hub at the IIT Ropar in the framework of the National Mission on Interdisciplinary Cyber-Physical Systems.
%
\bibliographystyle{ACM-Reference-Format}
\bibliography{main}

\end{document}